\documentstyle[preprint,aps,pra]{revtex}
\draft
\begin{document}
\tightenlines

\title{Two-fluid hydrodynamics of a Bose gas including damping from normal
fluid transport coefficients}

\author{T. Nikuni}
\address{Department of Physics, Tokyo Institute of Technology, Oh-okayama,
Meguro, Tokyo 152, Japan} 

\author{A. Griffin\cite{griffin} and E. Zaremba\cite{zaremba}}
\address{JILA, University of Colorado, Boulder, Colorado 80309-0440}

\date{\today}
\maketitle

\begin{abstract}
We extend our recent work on the two-fluid hydrodynamics of the
condensate and non-condensate in a trapped Bose gas by including 
the dissipation associated with viscosity and thermal conduction.
For purposes of illustration, we consider the hydrodynamic modes in the
case of a uniform Bose gas. A finite thermal conductivity and shear
viscosity give rise to a damping of the first and second sound modes 
in addition to that found previously due to the lack of diffusive
equilibrium between the condensate and non-condensate. The relaxational
mode associated with this equilibration process is strongly coupled to
thermal fluctuations and reduces to the usual thermal diffusion mode
above the Bose-Einstein transition. In contrast to the standard 
Landau two-fluid hydrodynamics, we predict a damped mode centered at 
zero frequency, in addition to the usual second sound doublet.

\end{abstract}
\pacs{PACS numbers: 03.75.Fi, 05.30Jp, 67.40.Db }

\section{introduction}
Since the original discovery in 1995, the subject of Bose-Einstein 
condensation in trapped atomic gases
has become a major field of research~\cite{RMP,varenna}.
Even though these systems consist of a very dilute vapor, they 
exhibit robust collective oscillations which are strongly influenced by
mean-field interactions and collisions between the atoms. 
There has been considerable theoretical work devoted to describing the
collective modes of a trapped gas at very low temperatures in terms of
the solution of the time-dependent Gross-Pitaevskii (GP) equation for
the macroscopic wavefunction of the condensate. As discussed in several
recent reviews \cite{RMP,varenna}, there is excellent agreement
between experimental observations for $T\ll T_{\rm BEC}$ and 
theoretical calculations based on the $T=0$ GP equation.

At elevated temperatures where the condensate is appreciably depleted by
thermal excitations, one enters a more complex regime in which 
collisions between the atoms must be considered. Two limiting cases for
the dynamics of the gas correspond to the collisionless and hydrodynamic
regimes~\cite{varenna,NP}.  In the collisionless regime, the main 
effect of the non-condensate component appears to be a shift in the
collective mode frequencies as a result of the change in the condensate
{\it number}, and to the appearance of Landau damping. There have been 
several studies of this collisionless region in trapped gases
where dynamic mean fields dominate the physics (we refer to the review 
articles in Ref.~\cite{varenna}).  The second regime arises when 
collisions between atoms are rapid enough to establish a state of 
dynamic local equilibrium in the non-condensate gas. This
collision-dominated hydrodynamic regime is the subject of this paper.
To be in this regime the collective modes of frequency $\omega$ must 
satisfy the condition $\omega\tau \ll 1$, where $\tau$ is some 
appropriate relaxation time for reaching local equilibrium.
As a rough estimate, we can take $\tau$ to be the collision time 
$\tau_{\rm cl}$ for a classical gas. Here 
$1/\tau_{\rm cl}=\sqrt{2}n\sigma \bar v$, where $\sigma=8\pi a^2$ is 
the low energy quantum mechanical cross-section for bosonic atoms
and $\bar v$ is the average thermal velocity~\cite{NZG}.
It is clear that a high density and/or a large atomic scattering 
cross-section are favorable for reaching the hydrodynamic region.
Experiments which can probe this region in trapped Bose gases are now
feasible and promise to provide much new physics.

One finds in the hydrodynamic regime that the dynamics of the condensate
and non-condensate components can {\it both} be described in terms of 
a few macroscopic (coarse-grained) variables (such as the local 
densities and velocities of the two components).
The coupled equations of motion for these local quantities will be referred to as the
two-fluid hydrodynamic equations.
The microscopic basis of these two-fluid equations rests on a generalized Gross-Pitaevskii
equation for the condensate atoms and a quantum kinetic equation for the non-condensate atoms.
These two components are coupled through mean-field interactions as well
as collisions between the atoms.
The authors have recently given a detailed derivation and discussion of such a two-fluid
hydrodynamics for trapped atomic gases at finite temperatures~\cite{ZGN,NZG,ZNG}.

These equations were derived for temperatures where the dominant thermal excitations
in the trap can be treated as atoms moving in a self-consistent Hartree-Fock field.
In particular, the original ZGN hydrodynamic equations derived in Ref.~\cite{ZGN} were
generalized in
Refs.~\cite{NZG,ZNG} to include collisions between condensate and non-condensate atoms.
This allows for the possibility of treating the situation in which
atoms of the non-condensate are in local thermodynamic equilibrium 
among themselves but are {\it not}
in diffusive equilibrium with the condensate atoms.
The resulting ZGN$'$ hydrodynamic equations~\cite{NZG,ZNG} involve a
characteristic relaxation time $\tau_{\mu}$ which is the time scale on
which local diffusive equilibrium is established. This equilibration
process leads to a novel damping mechanism which is associated with
the collisional exchange of atoms between the two components.
The ZGN$'$ equations are briefly reviewed in Section II.

In Section III, we further generalize the ZGN$'$ equations by
considering the effects of deviations from local equilibrium within 
the non-condensate.
At the finite temperatures of interest, this deviation from local equilibrium gives rise
to damping associated with the thermal conductivity and the shear viscosity.
This generalization has already been discussed in Section V of Ref.~\cite{NG} starting from
the ZGN hydrodynamic equations.

In the limit that the two components are in complete local equilibrium with each other,
our two-fluid hydrodynamic equations reduce to those first derived by Landau in
1941~\cite{Landau}.
The Landau two-fluid equations give an excellent description of the low frequency
response of superfluid $^4$He~\cite{Khal}.
As noted by several authors, the Landau theory is also valid for Bose-condensed gases.
The thermal conductivity and shear viscosity were first derived for a
uniform Bose-condensed gas at finite temperatures in a pioneering paper by Kirkpatrick and
Dorfman~\cite{KD}.
Their results were used by Gay and Griffin~\cite{GG} to evaluate the temperature-dependent
damping of first and second sound as predicted by the Landau two-fluid hydrodynamic equations.
Our present results are consistent with both of these early papers in the appropriate
Landau limit, namely when $\omega\tau_{\mu}\ll 1$\cite{NZG,ZNG}.
However it is important to point out that our generalized two-fluid
hydrodynamic equations provide a more complete description than the 
original Landau version since they can be used in situations in which 
the superfluid and normal fluid
are {\it not} in local diffusive equilibrium with each other.

To illustrate the physics, we use our ZGN$'$ equations in Section IV 
to study the hydrodynamic normal mode spectrum of a uniform Bose gas
in the presence of hydrodynamic dissipation. In particular, we
show how first and second sound modes are affected by viscosity and
thermal conduction, and also discuss how the new relaxational mode 
exhibited in Refs.~\cite{NZG,ZNG} is modified. In another paper, we 
apply these same equations to a discussion of the damping of the 
out-of-phase dipole mode recently observed~\cite{ZGN,MIT} in a 
trapped Bose gas.

\section{A REVIEW OF THE ZGN$'$ equations}

In this section, we first briefly review the finite temperature ZGN$'$ equations
\cite{ZGN,ZNG} based on the assumption that the non-condensate is in local
equilibrium.
In the next section, we calculate the corrections to these equations which arise
from a small deviation from local equilibrium.
The non-condensate atoms are described by the distribution function
$f({\bf r},{\bf p},t)$, which obeys the quantum kinetic equation
(we set $\hbar=1$ throughout this paper):
\begin{eqnarray}
{\partial f({\bf r},{\bf p},t) \over \partial t} + {{\bf p} \over m} 
\cdot \bbox{\nabla} f({\bf r},{\bf p},t) &-& \bbox{\nabla} U \cdot
\bbox{\nabla}_{{\bf p}} f({\bf r} ,{\bf p},t) \cr
&=& C_{12}[f] + C_{22}[f].
\label{eq1}
\end{eqnarray}
Here the effective potential
$U({\bf r},t)\equiv U_{\rm ext}({\bf r})+2g[n_c({\bf r},t)+\tilde 
n({\bf r},t)]$ includes the self-consistent Hartree-Fock (HF) mean field, and as usual,
we treat the inter-atomic interaction in the $s$-wave approximation with
$g=4\pi a/m$. 
The condensate density is $n_c({\bf r},t)\equiv|\Phi({\bf r},t)|^2$ 
and the non-condensate density $\tilde n({\bf r},t)$ is given by
\begin{equation}
\tilde n({\bf r},t)=\int\frac{d{\bf p}}{(2\pi)^3}f({\bf r},{\bf p},t).
\label{eq2}
\end{equation}
The two collision terms in (\ref{eq1}) are given by
\begin{eqnarray}
&&C_{22}[f] \equiv 4\pi g^2
 \int{d{\bf p}_2\over(2\pi)^3}\int{d{\bf p}_3\over(2\pi)^3}
\int d{\bf p}_4 \cr
&&\times\delta ({\bf p}+{\bf p}_2 -{\bf p}_3 -{\bf p}_4)
\delta(\tilde\varepsilon_{p}+\tilde\varepsilon_{p_2}
-\tilde\varepsilon_{p_3}-\tilde\varepsilon_{p_4}) \cr
&&\times\left[(1+f)(1+f_2)f_3f_4-ff_2(1+f_3)(1+f_4)\right]\, ,
\label{eq3}
\end{eqnarray}
\begin{eqnarray}
&&C_{12}[f]\equiv 4\pi g^2 n_c \int\frac{d{\bf p}_1}{(2\pi)^3}\int
d{\bf p}_2 \int d{\bf p}_3 \cr
&&\times  \delta(m{\bf v}_c+{\bf p}_1-{\bf p}_2-{\bf p}_3)
\delta(\varepsilon_c+\tilde\varepsilon_{p_1}
-\tilde\varepsilon_{p_2}-\tilde\varepsilon_{p_3}) \cr
&&\times  [\delta({\bf p}-{\bf p}_1)-\delta({\bf p}-{\bf p}_2)
-\delta({\bf p}-{\bf p}_3)] \cr
&&\times [(1+f_1)f_2f_3-f_1(1+f_2)(1+f_3)],
\label{eq4}
\end{eqnarray}
with $f \equiv f({\bf r, p}, t),\, f_i\equiv f({\bf r, p}_i, t)$.
The expression in (\ref{eq4}) takes into account the fact that a
condensate atom locally has energy
$\varepsilon_c({\bf r},t)=\mu_c({\bf r},t)+\frac{1}{2}mv_c^2({\bf r},t)$
 and momentum $m{\bf v}_c$, where the condensate chemical potential 
$\mu_c$ and velocity ${\bf v}_c$ will be defined shortly.
On the other hand, a non-condensate atom locally has the HF energy
$\tilde\varepsilon_p({\bf r},t)=\frac{p^2}{2m}+U({\bf r},t)$.
This particle-like dispersion relation limits our analysis 
to finite temperatures.

The equation of motion for the condensate was derived in Ref.~\cite{ZNG}
(see also Ref.~\cite{stoof}) and is given by a generalized 
Gross-Pitaevskii equation for the macroscopic wavefunction 
$\Phi({\bf r},t)$
\begin{equation}
i\frac{\partial\Phi({\bf r},t)}{\partial t}=\left[-\frac{\nabla^2}{2m}+U_{\rm ext}({\bf r})
+gn_c({\bf r},t)+2g\tilde n({\bf r},t)-iR({\bf r},t)\right]\Phi({\bf r},t),
\end{equation}
where
\begin{equation}
R({\bf r},t)=\frac{\Gamma_{12}({\bf r},t)}{2n_c({\bf r},t)},
\end{equation}
with
\begin{equation}
\Gamma_{12}\equiv\int\frac{d{\bf p}}{(2\pi)^3}C_{12}[f({\bf r},{\bf p},t)].
\end{equation}
The dissipative term $R$ in (5) is associated with the exchange of atoms
between the condensate and non-condensate as described by the collision
integral $C_{12}[f]$ in (4).
We see that (1) and (5) must be solved self-consistently.
It is customary to rewrite the GP equation (5) in terms of the amplitude and phase of
$\Phi({\bf r},t)=\sqrt{n_c({\bf r},t)}e^{i\theta({\bf r},t)}$, which leads to
(${\bf v}_c=\bbox{\nabla}\theta({\bf r},t)/m$)
\begin{eqnarray}
{\partial n_c \over \partial t} + \bbox{\nabla}\cdot(n_c{\bf v}_c)&=& 
-\Gamma_{12}[f]\,, \cr
m\left({\partial\over\partial t}+{\bf v}_c\cdot 
\bbox{\nabla}\right) {\bf v}_c&=&-\bbox{\nabla}\mu_c \ ,
\label{eq5b}
\end{eqnarray}
\label{eq5}

\noindent
where the condensate chemical potential is given by
\begin{equation}
\mu_c({\bf r}, t) =-\frac{\nabla^2\sqrt{n_c({\bf r},t)} }{2m\sqrt{n_c({\bf r},t)}}
+U_{\rm ext}({\bf r})+
gn_c({\bf r}, t)+2g\tilde{n}({\bf r}, t)\, .
\label{eq6}
\end{equation}
One sees that $\Gamma_{12}$ in (8) plays the role of a ``source function" in the 
continuity equation for the condensate, arising from the fact that $C_{12}$ collisions
do not conserve the number of condensate atoms \cite{ZNG}.
Because of the structure of the equations in (8), they are often referred to as
``hydrodynamic equations", even though they are completely 
equivalent to the generalized GP equation in (5).

Following the standard procedure in the classical kinetic theory of gases
\cite{huang}, we take moments
of (1) to derive the most general form of hydrodynamic equations for the
non-condensate. These moment equations take the form ($\mu$ and $\nu$ 
are Cartesian components):
\begin{eqnarray}
&&{\partial{\tilde n}\over\partial t}+\bbox{\nabla}\cdot 
(\tilde{n}{\bf v}_n) = \Gamma_{12}[f]\,, \cr
&&m{\tilde n}\left({\partial\over\partial t}+{\bf v}_n\cdot 
\bbox{\nabla}\right) v_{n\mu}=-{\partial P_{\mu\nu}\over\partial x_\nu}
-{\tilde n}{\partial U\over\partial x_\mu}
-m(v_{n\mu}-v_{c\mu})\Gamma_{12}[f]\,, \\
&&{\partial\tilde\epsilon\over\partial t} +
\nabla\cdot(\tilde\epsilon{\bf v}_n) = -\bbox{\nabla}\cdot{\bf Q}
-D_{\mu\nu} P_{\mu\nu}+ \left[\frac{1}{2}m({\bf v}_n-{\bf v}_c)^2
+\mu_c-U\right]\Gamma_{12}[f] \nonumber.
\end{eqnarray}
\label{eq33}

\noindent
The non-condensate density was defined earlier in (\ref{eq2}) while
the non-condensate local velocity is defined by
\begin{equation}
{\tilde n}({\bf r},t){\bf v}_n({\bf r}, t)\equiv\int{d{\bf p}
\over(2\pi)^3} {{\bf p}\over m} f({\bf r, p}, t)\,.
\label{eq34}
\end{equation}
In addition, we have

\begin{eqnarray}
P_{\mu\nu}({\bf r}, t)&\equiv& m
\int{d{\bf p}\over(2\pi)^3}\left({p_\mu\over m} - v_{n\mu}\right)
\left({p_\nu\over m} - v_{n\nu}\right)
f({\bf r, p}, t), \label{eq35a} \cr
{\bf Q}({\bf r}, t)&\equiv& \int{d{\bf p}\over(2\pi)^3} {1\over 2m} 
({\bf p}-m{\bf v}_n)^2\left({{\bf p}\over m}-{\bf v}_n\right)
f({\bf r, p}, t),\label{eq35b}  \\
\tilde \epsilon({\bf r}, t) &\equiv&\int{d{\bf p}\over(2\pi)^3}
{1\over 2m} ({\bf p}-m{\bf v}_n)^2
f({\bf r},  {\bf p}, t) \,. \nonumber
\end{eqnarray}
\label{eq35}
Finally, the symmetric rate-of-strain tensor appearing in (10)
is defined as
\begin{equation}
D_{\mu\nu}({\bf r}, t) \equiv {1 \over 2} \left({\partial v_{n \mu}
\over \partial x_\nu} + {\partial v_{n \nu} \over \partial x_\mu}
\right).
\label{eq36}
\end{equation}
Formally, these results are {\it exact} consequences of the kinetic equation (1).

The lowest order approximate solution of (1) is based on the assumption
that $C_{22}$ collisions are sufficiently rapid to force the
distribution function to have the form of the local equilibrium Bose 
distribution 
\begin{equation}
\tilde f({\bf r},{\bf p},t)=
\frac{1}{e^{\beta[\frac{1}{2m}({\bf p}-m{\bf v}_n)^2+U-\tilde \mu]}-1}
\, .
\label{eq8}
\end{equation}
Here, the temperature parameter $\beta$, normal fluid velocity 
${\bf v}_n$, chemical potential $\tilde \mu$, and mean field $U$ are all
functions of ${\bf r}$ and $t$. 
One may immediately verify that $\tilde f$ satisfies $C_{22}[\tilde f]=0$
independent of the value of $\tilde \mu$.  In contrast, 
one finds that $C_{12}[\tilde f]$ is in general finite, namely
\begin{eqnarray}
C_{12}[\tilde f]&=&4\pi g^2 n_c 
[1 - e^{-\beta(\tilde \mu-\frac{1}{2}m({\bf v}_n-{\bf v}_c)^2-\mu_c)}] 
\cr && \times \int\frac{d{\bf p}_1}{(2\pi)^3}
\int d{\bf p}_2
\int d{\bf p}_3
\delta(m{\bf v}_c+{\bf p}_1-{\bf p}_2-{\bf p}_3)
\delta(\tilde\varepsilon_1+\varepsilon_c
-\tilde\varepsilon_2-\tilde\varepsilon_3) \cr
&& \times [\delta({\bf p}-{\bf p}_1)-\delta({\bf p}-{\bf p}_2)
-\delta({\bf p}-{\bf p}_3)]
(1+\tilde f_1)\tilde f_2 \tilde f_3.
\label{eq40}
\end{eqnarray}
Using the local distribution function (14) 
to evaluate the moments in (2) and (12), we find that 
the heat current ${\bf Q}({\bf r},t) = 0$, and that
\begin{equation}
\tilde n({\bf r},t) = \int{d{\bf p}\over (2\pi)^3}
\left. \tilde f({\bf r, p}, t)\right|_{{\bf v}_n=0}
= {1\over\Lambda^3} g_{3/2}(z)\,,
\label{eq46}
\end{equation}
\begin{equation}
P_{\mu\nu} ({\bf r}, t) = \delta_{\mu\nu}{\tilde P}({\bf r}, t)
\equiv\delta_{\mu\nu}\int{d{\bf p}\over (2\pi)^3} {p^2\over 3m}
\left. \tilde f({\bf r, p}, t)\right|_{{\bf v}_n=0}
=\delta_{\mu\nu}\frac{1}{\beta\Lambda^3}g_{5/2}(z).
\label{eq41}
\end{equation}
Here $z({\bf r},t)\equiv e^{\beta[\tilde\mu-U({\bf r},t)]}$ is the local
fugacity, $\Lambda({\bf r},t) \equiv[{2\pi/ mk_B T({\bf r},t)}]^{1/2}$
is the local thermal de Broglie wavelength and 
$g_n(z)=\sum_{l=1}^{\infty}z^l/l^n$ are the Bose-Einstein functions.
The kinetic energy density  is given by
$\tilde \epsilon({\bf r},t) ={3\over 2} \tilde P ({\bf r},t)$ which is
the same relation as found for a uniform ideal gas. 

To summarize, using $f \simeq \tilde f$, we obtain the ZGN$'$ 
lowest-order hydrodynamic equations for the non-condensate given in 
Refs.~\cite{NZG,ZNG}
\begin{eqnarray}
&&{\partial{\tilde n}\over\partial t}+\bbox{\nabla}\cdot 
(\tilde{n}{\bf v}_n) = \Gamma_{12}[\tilde f]\,, \cr
&&m{\tilde n}\left({\partial\over\partial t}+{\bf v}_n\cdot 
\bbox{\nabla}\right) {\bf v}_n=-\bbox{\nabla} \tilde P
-{\tilde n}\bbox{\nabla} U
-m(v_{n\mu}-v_{c\mu})\Gamma_{12}[\tilde f]\,, \\
&&{\partial\tilde P\over\partial t} +
\nabla\cdot(\tilde P{\bf v}_n) = -{2\over 3}\tilde P \bbox{\nabla}
\cdot{\bf v}_n + {2\over 3} \left[\frac{1}{2}m({\bf v}_n-{\bf v}_c)^2
+\mu_c-U\right]\Gamma_{12}[\tilde f] \nonumber.
\end{eqnarray}
\label{eq46'}
where $\Gamma_{12}[\tilde f]$ is obtained from (7) 
with $C_{12}[\tilde f]$ given by (\ref{eq40}). 

\section{ZGN$'$ equations with hydrodynamic dissipation}

We next derive the additional terms which arise from the equations
in (10) due to a deviation of the distribution function from 
local equilibrium, $f\simeq \tilde f + \delta f$~\cite{huang}.
Following Refs.~\cite{KD,NG}, we write this deviation in the form
\begin{equation}
\delta f=\tilde f({\bf r},{\bf p},t) [1+\tilde f({\bf r},{\bf p},t)]
\psi({\bf r},{\bf p},t).
\end{equation}
To first order in $\psi$, the $C_{22}$ collision integral in (3) reduces
to
\begin{eqnarray}
C_{22}[\tilde f+\delta f]&\simeq& 4\pi g^2\int\frac{d{\bf p}_2}
{(2\pi)^3} \int\frac{d{\bf p}_3}{(2\pi)^3}\int d{\bf p}_4
\delta({\bf p}+{\bf p}_2-{\bf p}_3-{\bf p}_4)\delta(\tilde\varepsilon_1
+\tilde\varepsilon_2-\tilde\varepsilon_3-\tilde\varepsilon_4) \cr
&& \times \tilde f \tilde f_2(1+\tilde f_3)(1+\tilde f_4)
(\psi_3+\psi_4-\psi_2-\psi)\equiv \hat L_{22}[\psi].
\end{eqnarray}
In the left hand side of (1) and in the $C_{12}$ collision integral,
we approximate $f$ by the local Bose distribution $\tilde f$.
The various derivatives of ${\bf v}_n$, $\tilde\mu$, $T$ and $U$ with 
respect to ${\bf r}$ and $t$ can be written using the lowest-order 
hydrodynamic equations given in (18). 
The resulting linearized equation which determines $\psi$ is 
(for details, see the Appendix)
\begin{eqnarray}
&&\left\{\frac{{\bf u}\cdot\bbox{\nabla}T}{T}\left[\frac{mu^2}{2k_{\rm B}T}-
\frac{5g_{5/2}(z)}{2g_{3/2}(z)}\right]+\frac{m}{k_{\rm B}T}D_{\mu\nu}
\left(u_{\mu}u_{\nu}-\frac{1}{3}\delta_{\mu\nu}u^2\right) \right. \cr
&&
+\left.\left(\sigma_2+\frac{mu^2}{3k_{\rm B}T}\sigma_1+\frac{m}{k_{\rm B}T}
{\bf u}\cdot{\bf w} \right)\frac{\Gamma_{12}[\tilde f]}{\tilde n}\right\}\tilde f(1+\tilde f)
-C_{12}[\tilde f]=\hat L_{22}[\psi].
\end{eqnarray}
Here the thermal velocity ${\bf u}$ is defined by 
$m{\bf u}\equiv {\bf p}-m{\bf v}_n$ and ${\bf w}\equiv {\bf v}_c-{\bf v}_n$.
The dimensionless thermodynamic functions $\sigma_1$, $\sigma_2$ are 
defined by
\begin{eqnarray}
\sigma_1({\bf r},t)&\equiv&\frac{\gamma\tilde n
\left[\frac{1}{2}mw^2+\mu_c-U\right]-\frac{3}{2}\tilde n^2}
{\frac{5}{2}\tilde P\gamma-\frac{3}{2}\tilde n^2}, \cr
\sigma_2({\bf r},t)&\equiv&\beta\frac{\frac{5}{2}\tilde P\tilde n
-\tilde n^2 \left[\frac{1}{2}mw^2+\mu_c-U\right]}
{\frac{5}{2}\tilde P\gamma-\frac{3}{2}\tilde n^2},
\end{eqnarray}
where $\gamma({\bf r},t)\equiv \frac{\beta}{\Lambda^3}g_{1/2}(z({\bf r},t))$.
We note that Refs.~\cite{NZG,ZNG} introduce a related dimensionless quantity
$\tilde\gamma\equiv g\gamma$.

Since (21) is a linear equation for $\psi$, one may write the solution as 
$\psi=\psi^{(1)}+\psi^{(2)}$, where $\psi^{(1)}$ is the solution of 
\begin{equation}
\left\{\frac{{\bf u}\cdot\bbox{\nabla}T}{T}\left[\frac{mu^2}{2k_{\rm B}T}-
\frac{5g_{5/2}(z)}{2g_{3/2}(z)}\right]+\frac{m}{k_{\rm B}T}D_{\mu\nu}
\left(u_{\mu}u_{\nu}-\frac{1}{3}\delta_{\mu\nu}u^2\right)\right\}
\tilde f(1+\tilde f)=\hat L_{22}[\psi^{(1)}],
\end{equation}
and $\psi^{(2)}$ is the solution of
\begin{equation}
\left(\sigma_2+\frac{mu^2}{3k_{\rm B}T}\sigma_1+\frac{m}{k_{\rm B}T}
{\bf u}\cdot{\bf w} \right)\frac{\Gamma_{12}[\tilde f]}{\tilde n}\tilde f(1+\tilde f)
-C_{12}[\tilde f]=\hat L_{22}[\psi^{(2)}].
\end{equation}
We note from its definition that $\tilde f({\bf r},{\bf p},t)$ is a
function of the variable $u^2$ while the linearized operator 
$\hat L_{22}$ defined in (20) is a function of ${\bf u}$.
The expression (15) for $C_{12}[\tilde f]$ can be written in the form
\begin{eqnarray}
C_{12}[\tilde f]&=&\frac{2g^2n_cm^3}{(2\pi)^2}
[1-e^{-\beta(\mu_{\rm diff}-\frac{1}{2}mw^2)}]
\int d{\bf u}_1\int d{\bf u}_2 \int d{\bf u}_3 \cr
&&\times\delta({\bf w}+{\bf u}_1
-{\bf u}_2-{\bf u}_3)
\delta(\mu_c-U+{\small \frac{m}{2}}(w^2+u_1^2-u_2^2-u_3^2)) \cr
&&\times [\delta({\bf u}-{\bf u}_1)-\delta({\bf u}-{\bf u}_2)
-\delta({\bf u}-{\bf u}_3)](1+\tilde f_1)\tilde f_2 \tilde f_3,
\end{eqnarray}
where $\mu_{\rm diff}\equiv\tilde \mu-\mu_c$.
This expression shows that $C_{12}[\tilde f]$ depends on ${\bf u}$
and ${\bf w}$, and in particular, obeys the relation $C_{12}[{\bf u},{\bf w}]
=C_{12}[-{\bf u},-{\bf w}]$.
Finally, we note that $\Gamma_{12}[\tilde f]$ in (24) is independent of ${\bf u}$, but
depends on ${\bf w}$.

The equations (23) and (24) can be shown to have unique solutions if we
impose the following constraints on $\psi^{(1)}$ and $\psi^{(2)}$:
\begin{equation}
\int d{\bf p} \ \tilde f(1+\tilde f)\psi^{(i)} =\int d{\bf p}
 \ p_{\mu}\tilde f(1+\tilde f)\psi^{(i)}
=\int d{\bf p} \ p^2\tilde f(1+\tilde f)\psi^{(i)} =0.
\end{equation}
Physically these constraints mean that the deviations from local
equilibrium make no contribution to $\tilde n$, ${\bf v}_n$ and the 
diagonal component of $P_{\mu\nu}$. The solution for $\psi^{(1)}$
of (23) has been given already in Ref.~\cite{NG}, namely
\begin{equation}
\psi^{(1)}=\left[\frac{\bbox{\nabla}T\cdot{\bf u}}{T}A(u)+2D_{\mu\nu}
\left(u_{\mu}u_{\nu}-\frac{1}{3}u^2\delta_{\mu\nu}\right)B(u)\right].
\end{equation}
Using this solution for $\psi^{(1)}$, one finds that the heat current 
density ${\bf Q}$ and the pressure tensor $P_{\mu\nu}$ are given by
\begin{eqnarray}
P_{\mu\nu}&=&\delta_{\mu\nu} \tilde P-2\eta\left[D_{\mu\nu}-\frac{1}{3}
{\rm Tr}D \delta_{\mu\nu}\right]+P^{(2)}_{\mu\nu}, \cr
{\bf Q}&=&-\kappa\bbox{\nabla}T+{\bf Q}^{(2)},
\end{eqnarray}
where $P^{(2)}_{\mu\nu}$ and ${\bf Q}^{(2)}$ are the contribution from 
$\psi^{(2)}$. The explicit expressions for the transport coefficients 
$\eta$ and $\kappa$ are associated with $\psi^{(1)}$. They are given in
Eqs.~(38) and (34), respectively, of Ref.~\cite{NG} for a trapped Bose
gas below $T_{\rm BEC}$.
The analogous transport coefficients for a uniform degenerate Bose gas 
above $T_{\rm BEC}$ were first calculated by Uehling and 
Uhlenbeck~\cite{UU}.

The additional corrections due to $\psi^{(2)}$ are given by
\begin{eqnarray}
P_{\mu\nu}^{(2)}&=&m\int \frac{d{\bf p}}{(2\pi)^3}u_{\mu}u_{\nu}
\tilde f(1+\tilde f)\psi^{(2)}, \cr
{\bf Q}^{(2)}&=&\int \frac{d{\bf p}}{(2\pi)^3}\frac{m}{2}u^2{\bf u}
\tilde f(1+\tilde f)\psi^{(2)}.
\end{eqnarray}
The most general solution of (24) for $\psi^{(2)}$ is of the form
\begin{equation}
\psi^{(2)}=[1-e^{-\beta(\mu_{\rm diff}-\frac{1}{2}mw^2)}]D({\bf u},{\bf w}).
\end{equation}
If ${\bf w}\equiv{\bf v}_n-{\bf v}_c=0$, the left hand side of (24) and 
$D({\bf u})$ are isotropic functions of ${\bf u}$.
Using this fact in conjunction with the relation
 $C_{12}[{\bf u},{\bf w}]=C_{12}[-{\bf u},-{\bf w}]$ as
noted below (25), we conclude that $D$ must have the following form
\begin{equation}
D({\bf u},{\bf w})\simeq D_0(u)+D_1(u){\bf w}\cdot{\bf u}
+D_2(u)w_{\mu}w_{\nu}\left(u_{\mu}u_{\nu}-\frac{1}{3}u^2\delta_{\mu\nu}\right)
+O(w^3).
\end{equation}
Using this, the additional terms given by (29) have the following form
\begin{eqnarray}
P^{(2)}_{\mu\nu}&\propto&[1-e^{-\beta(\mu_{\rm diff}-\frac{1}{2}mw^2)}]
\left(w_{\mu}w_{\nu}-\frac{1}{3}w^2\delta_{\mu\nu}\right), \cr
{\bf Q}^{(2)}&\propto& [1-e^{-\beta(\mu_{\rm diff}-\frac{1}{2}mw^2)}]
{\bf w}.
\end{eqnarray}
We note that the constraints in (26) imply that
\begin{equation}
\int d{\bf u}\ u^2\tilde f(1+\tilde f)\psi^{(i)}=0,
\end{equation}
and thus it follows that the isotropic term $D_0(u)$ in (31) makes no 
contribution to $P^{(2)}_{\mu\nu}$ in (29).

In summary, we have obtained the following hydrodynamic equations for 
the non-condensate including the normal fluid transport coefficients
\label{eq26}
\begin{eqnarray}&&\frac{\partial \tilde n}{\partial t}
+\bbox{\nabla}\cdot(\tilde n{\bf v}_n)=\Gamma_{12}[\tilde f],
\label{eq26a}\cr
&&mn\left(\frac{\partial}{\partial t}+{\bf v}_n\cdot \bbox{\nabla}\right)v_{n\mu}
+\frac{\partial \tilde P}{\partial x_{\mu}}+\tilde n
\frac{\partial U}{\partial
x_{\mu}}=-m(v_{n\mu}-v_{c\mu})\Gamma_{12}[\tilde f] \cr
&& \ \ \ \ \ \ \ \ \ \ \ \ \ \ \ \ \ \ \ \ \ \ \ \ 
+\frac{\partial}{\partial x_{\nu}}\left\{2\eta\left[D_{\mu\nu}-\frac{1}{3}
({\rm Tr}D)\delta_{\mu\nu} \right]\right\}
-\frac{\partial P^{(2)}_{\mu\nu}}{\partial x_{\nu}},
\label{eq26b}\cr
&&\frac{\partial \tilde \epsilon}{\partial t}+\bbox{\nabla}
\cdot(\tilde\epsilon{\bf v}_n)
+(\bbox{\nabla}\cdot{\bf v}_n)\tilde P=\left[\frac{1}{2}m({\bf v}_n-{\bf v}_c)^2
\mu_c-U\right]\Gamma_{12}[\tilde f] \cr
&& \ \ \ \ \ \ \ \ \ \ \ \ \ \ \ \ \ \ \ \ \ \ \ \ \ \ 
+\bbox{\nabla}\cdot(\kappa\bbox{\nabla} T)
+2\eta\left[ D_{\mu\nu}-\frac{1}{3}({\rm Tr}D)\delta_{\mu\nu}\right]^2
-\bbox{\nabla}{\bf Q}^{(2)}.
\label{eq26c}
\end{eqnarray}
Since $P_{\mu\nu}^{(2)}$ and ${\bf Q}^{(2)}$ in (32) are at least of 
second order in the fluctuations in $\delta\mu_{\rm diff}$ and 
$\delta{\bf w}$ around static equilibrium, these terms can be neglected
when discussing the {\it linearized} form of these hydrodynamic 
equations.  The equivalent ``quantum"
hydrodynamic equations for the condensate are given in (8).
 
In closing this Section, it is useful to summarize the logical structure of our analysis.
The kinetic equation (1) leads to the exact set of equations involving
the variables $\tilde n$, ${\bf v}_n$, $\tilde\varepsilon$, 
$P_{\mu\nu}$ and ${\bf Q}$ which are defined in terms of various 
moments of the distribution function $f({\bf r},{\bf p},t)$. In
addition, the equations in (10) contain the function $\Gamma_{12}$ in 
(7) associated with collisions between condensate and non-condensate 
atoms.  In Refs.~\cite{NZG,ZNG}, the closed set of hydrodynamic 
equations displayed in (18) were derived by making use of
the local equilibrium distribution function. In the present paper, 
we have extended this analysis to include a small deviation (19)
from local equilibrium, following the Chapman-Enskog approach.
However, we have only included contributions to the function $\psi({\bf r},{\bf p},t)$
[see (19)] associated with the linearized $C_{22}$ collision integral
($\hat L_{22}$ in (20)), which leads to the linear integral equations for $\psi$ in (23)
and (24).
We have omitted any contribution to $\psi({\bf r},{\bf p},t)$ associated with the linearized
$C_{12}$ collision integral.

Our neglect of the latter contribution can be justified by noting that
$C_{12}$ in (4) is proportional to the condensate density 
$n_c({\bf r},t)$. Thus the deviations from local equilibrium due to 
the $C_{12}$ collision integral are relatively unimportant 
at temperatures close to $T_{\rm BEC}$.
However at lower temperatures, the deviations from local equilibrium due
to $C_{12}$ collisions will become increasingly important and one can 
expect corrections to the values of $\kappa$, $\eta$ and $\tau_{\mu}$ 
obtained in the present paper.  Such corrections were evaluated in 
Ref.~\cite{KD} for a uniform Bose-condensed gas, although these authors
made the further assumption that the condensate and non-condensate were 
in diffusive equilibrium with each other (i.e., $\tau_{\mu}\to 0$).
As expected, these corrections to the values of $\kappa$ and $\eta$
are of order $n_c/\tilde n$ at temperatures close to $T_{\rm BEC}$
(see Eq.~(26) of the second reference in~\cite{KD}) and hence can be 
neglected. 
 
\section{normal modes for a uniform Bose gas}
In this section, we discuss the normal mode solutions of the linearized
hydrodynamic equations, as given by
\begin{eqnarray}
\frac{\partial\delta n_c}{\partial t}&=&-\bbox{\nabla}\cdot(n_{c0}\delta{\bf v}_c)
-\delta\Gamma_{12}, \cr
\frac{\partial\delta{\bf v}_c}{\partial t}&=&-\bbox{\nabla}\delta\mu_c,
\end{eqnarray}
\label{eq42}
and
\begin{eqnarray}
&&\frac{\partial \delta \tilde n}{\partial t}+\bbox{\nabla}
 \cdot(\tilde n_0\delta{\bf v}_n)=\delta\Gamma_{12}, 
\label{eq42a} \cr
&&m\tilde n_0\frac{\partial \delta v_{n\mu}}{\partial t}
=-\frac{\partial \delta \tilde P}{\partial x_{\mu}} 
- \delta \tilde n \frac{\partial U}{\partial x_{\mu}} 
-2g\tilde n_0\frac{\partial \delta n}{\partial x_{\mu}} +
\frac{\partial}{\partial x_{\nu}}\left\{2\eta\left[D_{\mu\nu}-\frac{1}{3}
({\rm Tr}D)\delta_{\mu\nu} \right]\right\},
\label{eq42b} \cr
&&\frac{\partial \delta \tilde P}{\partial t}=
-\frac{5}{3}\bbox{\nabla}\cdot(\tilde P_0\delta {\bf v}_n)
+\frac{2}{3}\delta{\bf v}_n\cdot\bbox{\nabla} \tilde P_0
+(\mu_{c0}-U_0)\delta\Gamma_{12}
+\frac{2}{3}\bbox{\nabla}\cdot(\kappa\bbox{\nabla} \delta T).
\label{eq42c}
\end{eqnarray}
As discussed in Refs.~\cite{NZG,ZNG}, 
the source term $\delta \Gamma_{12}$ is conveniently expressed in terms of the
fluctuation of the local chemical potential difference 
$\mu_{\rm diff}\equiv\tilde\mu-\mu_c$,
\begin{equation}
\delta\Gamma_{12}=-\frac{\beta_0n_{c0}}{\tau_{12}}\delta\mu_{\rm diff}.
\end{equation}
Here we have introduced an equilibrium
relaxation time involving collisions between
condensate and non-condensate atoms,
\begin{eqnarray}
&&\frac{1}{\tau_{12}}\equiv4\pi g^2
\int\frac{d{\bf p}_1}{(2\pi)^3}\int \frac{d{\bf p}_2}{(2\pi)^3}
\int d{\bf p}_3 (1+\tilde f_1)\tilde f_2 \tilde f_3 \cr
&& \ \ \times\delta({\bf p}_1-{\bf p}_2-{\bf p}_3)
\delta(\mu_{c}+\tilde\varepsilon_{p_1}
-\tilde\varepsilon_{p_2}-\tilde\varepsilon_{p_3}),
\end{eqnarray}
where it is understood that all quantities now pertain to static 
thermal equilibrium.
 
One can now eliminate $\delta T$ and $\delta \mu_{\rm diff}$ from the above linearized 
equations using the relations (see (A6) and (A4) of Appendix A)
\begin{eqnarray}
\delta \tilde\mu&=&\frac{\sigma_{10}}{\tilde n_0}\delta \tilde P
+\frac{\sigma_{20}}{\beta_0\tilde n_0}\delta \tilde n+2g(\delta\tilde n+\delta n_c), \cr
\delta T&=&\frac{T_0\sigma_{30}}{\tilde P_0}\delta \tilde P
-\frac{T_0\sigma_{40}}{\tilde n_0}\delta \tilde n.
\end{eqnarray}
The various coefficients appearing are given by
\begin{eqnarray}
\sigma_{10}&=&-\frac{\frac{3}{2}\tilde n_0^2-\gamma_0\tilde n_0(\mu_{c0}-U_0)}
{\frac{5}{2}\tilde P_0\gamma_0-\frac{3}{2}\tilde n_0^2}, \ \  
\sigma_{20}=\beta_0\tilde n_0
\frac{\frac{5}{2}\tilde P_0-\tilde n_0(\mu_{c0}-U_0)}
{\frac{5}{2}\tilde P_0\gamma_0-\frac{3}{2}\tilde n_0^2}, \cr
\sigma_{30}&=&\frac{\tilde P_0 \gamma_0}
{\frac{5}{2}\tilde P_0\gamma_0-\frac{3}{2}\tilde n_0^2}, \ \ 
\sigma_{40}=\frac{\tilde n_0^2}
{\frac{5}{2}\tilde P_0\gamma_0-\frac{3}{2}\tilde n_0^2},
\end{eqnarray}
where $\mu_{c0}-U_0=-gn_{c0}$ in the Thomas-Fermi 
approximation~\cite{RMP}. Using (39) and (40), we see that our two-fluid
hydrodynamic equations in (35) and (36) reduce to five coupled equations
(three for the non-condensate and two for the condensate)
in the five variables $\delta \tilde n$, $\delta n_c$, 
$\delta{\bf v}_n$, $\delta {\bf v}_c$, and $\delta \tilde P$. Thus one 
expects a total of five normal modes from the extended ZGN$'$ equations.

We now consider the special case of a {\it uniform} gas 
$(U_{\rm ext}=0)$ and look for plane-wave solutions $\sim 
e^{i({\bf k}\cdot{\bf r}-\omega t)}$.
It is convenient to introduce dimensionless variables 
\begin{eqnarray}
\bar{n}_c\equiv n_c/n, &\quad& \bar{n}\equiv \tilde n/n, \cr
\delta \bar{v}_c\equiv i\hat{\bf k}\cdot\delta{\bf v}_c/v_{\rm cl},
&\quad& 
\delta \bar{v}_n\equiv i\hat{\bf k}\cdot\delta{\bf v}_n/v_{\rm cl}, \cr
\bar{P}\equiv \tilde P/nk_{\rm B}T_0, \qquad t&\equiv& T/T_{\rm BEC},
\qquad \lambda\equiv gn/k_{\rm B}T_{\rm BEC},
\end{eqnarray}
where $v_{\rm cl}\equiv (5k_{\rm B}T_{\rm BEC}/3m)^{1/2}$ is the sound 
velocity of a classical gas at $T=T_{\rm BEC}$.
We also introduce dimensionless frequency and wavenumber variables
\begin{equation}
\bar\omega\equiv\omega\tau_0,  \ \bar k\equiv kv_{\rm cl}\tau_0,
\end{equation}
where $\tau_0^{-1}\equiv\sigma n(16k_{\rm B}T_{\rm BEC}/\pi m)^{1/2}$ 
is the {\it classical} gas collision time \cite{NP},
evaluated at $T=T_{\rm BEC}$ (but with the quantum collision
cross-section for bosons, $\sigma=8\pi a^2$.) Finally we define 
dimensionless transport coefficients and a dimensionless collision time
\begin{equation}
\bar\kappa\equiv \kappa/nv_{\rm cl}^2\tau_0k_{\rm B}, \ 
\bar\eta\equiv\eta/nv_{\rm cl}^2m\tau_0, \ 
\bar{\tau}_{12}\equiv\tau_{12}/\tau_0.
\end{equation}
In terms of the above dimensionless 
quantities, we obtain the following closed set of equations 
\begin{mathletters}
\begin{eqnarray}
i\bar\omega\delta \bar{n}_c&=&
-\frac{\lambda\bar{n}_{c0}}{\bar{\tau}_{12} t}
\delta\bar{n}_c+\bar k\bar{n}_{c0}\delta\bar{v}_c-
\frac{\sigma_{10}\bar{n}_{c0}}{\bar{\tau}_{12}\bar{n}_0}
\delta \bar{P}
-\frac{\sigma_{20}\bar{n}_{c0}}{\bar{\tau}_{12}\bar{n}_0}\delta\bar{n},
\\
i\bar\omega\delta \bar{v}_c&=&-\frac{3}{5}\lambda \bar k(\delta\bar{n}_c
+2\delta\bar n), \\
i\bar\omega\delta\bar n&=& 
\frac{\lambda\bar{n}_{c0}}{\bar{\tau}_{12} t}
\delta \bar{n}_c
+\bar k \bar{n}_{0}\delta \bar{v}_n
+\frac{\sigma_{10}\bar{n}_{c0}}
{\bar{\tau}_{12}\bar{n}_0}\delta \bar P
+\frac{\sigma_{20}\bar{n}_{c0}}{\bar{\tau}_{12}\bar{n}_0}\delta \bar{n},
\\
i\bar\omega\delta \bar{v}_n&=&-\frac{6}{5}\lambda\bar k(\delta\bar{n}_c
+\delta \bar n)-\frac{3t}{5\bar{n}_0}\bar k\delta \bar{P}
+\frac{4\bar\eta}{3\bar{n}_0}\bar k^2 \delta\bar v_n,  \\
i\bar\omega\delta\bar{P}
&=&\frac{5\bar{P}_0}{3}\bar k\delta \bar{v}_n
-\frac{2}{3}\left(\frac{\sigma_{20}\lambda\bar{n}_{c0}^2}{\bar{\tau}_{12}t
\bar{n}_0}
+\bar k^2\frac{\sigma_{40}\bar\kappa}{\bar{n}_0}\right)\delta\bar{n}
-\frac{2\lambda^2\bar{n}_{c0}^2}{3\bar{\tau}_{12}t^2}\delta \bar{n}_c \cr
&&-\frac{2}{3}\left(\frac{\sigma_{10}\lambda\bar{n}_{c0}^2}{\bar{\tau}_{12}t
\bar{n}_0}-\bar k^2\frac{\sigma_{30}\bar\kappa}{\bar{P}_0}\right)
\delta \bar{P}.
\end{eqnarray}
\end{mathletters}
We emphasize that our hydrodynamic equations are restricted by the
assumption of local equilibrium and thus are only valid if 
$\bar\omega\ll 1$, $\bar k\ll 1$.  

It is straightforward to solve the coupled set of equations in (44).
Above $T_{\rm BEC}$, they reduce to three equations for three variables,
which gives rise to two
sound modes $(\pm uk)$ and a thermal diffusion mode.
Below $T_{\rm BEC}$, these equations give two first sound modes ($\pm u_1k$),
two second sound modes ($\pm u_2k$) and one relaxational mode.
If the transport coefficients $\eta$ and $\kappa$ 
are set to zero, these equations are identical to the
ZGN$'$ equations as discussed in Refs.~\cite{NZG,ZNG}.
Using (44), we can discuss the effect of $\eta$ and $\kappa$ on the
first and second sound modes, as well as on the relaxational mode.
In Fig.~1 of Ref.~\cite{ZNG}, a graph is given of the first and
second sound velocities vs. temperature. They are almost identical in 
both the ZGN limit $(\omega\tau_{\mu}\gg 1)$ and the Landau limit
($\omega\tau_{\mu}\ll 1$). Calculations based on the generalized 
ZGN$'$ equations in (44) show essentially no change in the first and 
second sound velocities when the effects of $\kappa$ and $\eta$ are 
included, for $\bar k < 0.5$. In Figs.~1--3, we  illustrate the effect 
of these transport coefficients on the damping $\Gamma$ of the first 
and second sound modes (where $\omega_i=uk_i-i\Gamma_i$) and on the
relaxational mode, all as a function of the dimensionless wavevector 
$\bar k$. These specific results are for a temperature of 
$T/T_{\rm BEC}=0.9$ and $gn/k_{\rm B}T_{\rm BEC}=0.2$, where one finds
$\tau_0/\tau_{\mu}=0.91$, $\bar\kappa=2.41$ and $\bar\eta=0.34$.
We note that in the absence of $\eta$ and $\kappa$, only second sound 
is significantly damped through coupling to the relaxational mode. When
we include $\eta$ and $\kappa$, $\Gamma_1$ is large compared to 
$\Gamma_2$.

Figs. 1--3 only show the damping in a case where $\omega\tau_{\mu}\ll 
1$. It is interesting to see how the relaxational mode depends on the 
value of $\omega\tau_{\mu}$.  In Fig.~1 of Ref.~\cite{NZG}, we have 
plotted the temperature dependence of $\tau_{12}$, $\tau_0$ and 
$\tau_{\mu}$ for $gn=0.1k_{\rm B}T_{\rm BEC}$. In Fig.~4, we plot the 
damping of the relaxational mode for $\bar k=0.4$, as a function of the
temperature. We also show the corresponding temperature dependence of 
$\tau_{\mu}/\tau_0$. As discussed in Ref.~\cite{NZG}, our use of a 
particle-like spectrum with a HF mean-field gives rise to a spurious 
finite value of the condensate density $n_{c0}$ at $T_{\rm BEC}$.
As a result, $\tau_{\mu}$ is also finite at $T_{\rm BEC}$.

In Appendix B, we derive an analytical expression for the width of the 
mode centered at zero frequency.
Working to first order in $\kappa$, $\eta$, $1/\tau_{12}$ and
second order in $\lambda=gn/k_{\rm B}T_{\rm BEC}$, we obtain
\begin{equation}
\Gamma_{\rm R}\simeq\frac{1}{\tau_{\mu}}+\frac{2}{5}
\frac{\sigma_{40}\kappa} {(\tilde P_0/T_0)}k^2.
\end{equation}
As shown in Fig.~4, this approximate expression is in good agreement 
with the result of a direct numerical evaluation of $\Gamma_{\rm R}$ 
from (44) at temperatures close to $T_{\rm BEC}$ (it deviates more and 
more at lower temperatures because of the non-linear
dependence of $\Gamma_{\rm R}$ on $\kappa$ and $1/\tau_{12}$).
The approximate analytic expression in (45) is useful in that it 
shows clearly that below $T_{\rm BEC}$, the relaxational mode is 
strongly coupled to thermal diffusion which arises from the deviation 
from local equilibrium of the non-condensate distribution function.
Indeed, one may view this damped mode as a renormalized version of the 
well-known thermal
diffusion mode found above $T_{\rm BEC}$~\cite{huang}.
In the classical high-temperature limit, we have $\sigma_{40}=1$ and
$\tilde P_0=nk_{\rm B}T_0$, and (45) reduces in this case
to $\Gamma_{\rm R}=\kappa k^2/nC_{\rm P}$, where the specific
heat $C_{\rm P}=5k_{\rm B}/2$. This is the well-known classical gas 
result for the thermal diffusion mode~\cite{GG,huang}. 

In the ZGN limit ($\tau_{\mu}\to\infty$), (45) shows that the damping
rate of the relaxational mode is mainly  due to the finite thermal 
conductivity (if we set $\kappa=0$, we obtain the zero frequency mode 
discussed in Section V of Ref.~\cite{ZNG}). In the opposite Landau 
limit of the ZGN$'$ equations ($\tau_{\mu}\to0$), the width of this
mode is dominated by $1/\tau_{12}$, associated with the $C_{12}$ 
collisions which bring about diffusive equilibrium between the 
condensate and non-condensate components (see Fig.~4). Eq.~(45) is in 
complete agreement with the qualitative picture sketched at the end of
Section V of Ref.~\cite{ZNG}. The mode spectrum which the coupled 
ZGN$'$ equations in (44) imply below $T_{\rm BEC}$ is different from 
the usual Landau two-fluid hydrodynamics~\cite{Khal,KD,NP}.
In the latter theory, the thermal diffusion mode above $T_{\rm BEC}$ 
does not persist as a relaxational mode below $T_{\rm BEC}$. Rather, 
the thermal diffusion mode is interpreted as being replaced by two 
damped second sound modes ($\pm u_2k-i\Gamma_2$) below $T_{\rm BEC}$.

Fig.~4 also shows that the width of the relaxational mode increases
sharply (over and above the spurious jump at $T_{\rm BEC}$ noted above)
as the temperature passes from above to below $T_{\rm BEC}$, as a 
result of the onset of $C_{12}$ collisions. This increased width of a 
mode peaked at $\omega=0$ will be a useful experimental signature of 
the new physics implied by the extended ZGN$'$ hydrodynamic equations
derived in the present paper. Of course, all of the above analysis says
nothing about the relative weight of the
first sound, second sound and relaxational modes below $T_{\rm BEC}$.
This requires the evaluation of a specific response function, such as 
the dynamic structure factor \cite{MIT2}.
Further calculations along these lines are in progress.

\section{conclusions}
In this paper, we have extended our recent derivation \cite{NZG,ZNG} of 
two-fluid hydrodynamic equations (referred to as the ZGN$'$ equations) 
to include the effects of a small deviation from local equilibrium of 
the non-condensate atoms. This brings in the usual kind of hydrodynamic
damping due to the thermal conductivity and shear viscosity of the 
thermal cloud. In calculating these transport coefficients we have only
taken into account the deviations from local equilibrium due to the 
$C_{22}$ collision integral but have neglected the contribution coming 
from the $C_{12}$ collision integral. This limits the validity of the 
present calculations to the
vicinity of the transition temperature. The damping due to hydrodynamic
dissipation is in addition to that due to the equilibration of the
condensate and non-condensate components already contained
in the original ZGN$'$ equations, as discussed in Refs.~\cite{NZG,ZNG}.

For illustration, we presented some numerical results for a {\it 
uniform} Bose gas. In this case, we find damped first and second sound 
modes ($\pm u_1k-i\Gamma_1, \pm u_2k-i\Gamma_2$) and a purely 
relaxational mode $\omega=-i\Gamma_{\rm R}$. The latter mode is not 
exhibited by the usual two-fluid hydrodynamic equations which assume
that the condensate and non-condensate are in local thermodynamic
equilibrium. The overall effect of $\eta$ and $\kappa$ on the ZGN$'$ 
predictions in Ref.~\cite{ZNG} is to introduce additional damping of 
all three modes. Above $T_{\rm BEC}$, the relaxational and second
sound modes $(\pm u_2k)$ merge into a single thermal diffusive mode.
The relaxational mode below $T_{\rm BEC}$ may therefore be viewed as the
superfluid analogue of the normal thermal diffusion mode.

In another paper~\cite{ZNG2}, we shall use these ZGN$'$ equations with 
transport coefficients to discuss the damping of hydrodynamic modes in 
a {\it trapped} Bose gas. In general this is a more complex situation, 
since one must contend with the fact that the local equilibrium 
solution becomes invalid as a starting point in the low density tail 
of the thermal cloud~\cite{NG,KPS}. However, one can derive general 
expressions for the damping and renormalization of the hydrodynamic 
modes as given by the ZGN equations. In particular, we shall show that 
the damping of the out-of-phase dipole oscillation of the condensate 
and non-condensate components in a trapped Bose-condensed gas
is only weakly affected by the non-condensate transport coefficients.
As a result, the damping of this mode is entirely due to the collisions
between the condensate and non-condensate atoms which are responsible 
for bringing
these two components into diffusive equilibrium. Further experimental 
studies~\cite{MIT} of this analogue of second sound in superfluid 
$^4$He~\cite{Khal} would be of great interest.
\begin{center}
{\bf ACKNOWLEDGMENTS}
\end{center}
We congratulate Prof. Boris Stoicheff for his many important
contributions to atomic, laser and condensed matter physics over a 
long and distinguished career in Canada. We are sure that if Boris 
were just a little younger, he would be doing research on BEC
in atomic gases and eager to work on the topics discussed in
this paper! A.G. would especially like to thank Boris for
many stimulating and informative discussions over the last 30 years 
and which still continue today.

A.G. and E.Z. would like to thank Eric Cornell and the BEC group at 
JILA for providing a stimulating environment and support during a 
six-month sabbatical in 1999 when this work was initiated. T.N. was 
supported by JSPS, while A.G. and E.Z. were supported by NSERC.

\begin{appendix}
\section{}

We briefly sketch the derivation of the kinetic equation in (21).
Using (14) in the left hand side of (\ref{eq1}), one has
\begin{eqnarray}
&&\left[ \frac{\partial}{\partial t} +\frac{{\bf p}}{m}\cdot
\bbox{\nabla} -\bbox{\nabla} U({\bf r},t)\cdot\bbox{\nabla}_{\bf p}\right]
\tilde f({\bf r},{\bf p},t) \nonumber \\
&&=\left[\frac{1}{z}\left(\frac{\partial}{\partial t}+
\frac{{\bf p}}{m}\cdot\bbox{\nabla} \right)z
+\frac{mu^2}{2k_{\rm B}T^2}
\left(\frac{\partial}{\partial t}+
\frac{{\bf p}}{m}\cdot\bbox{\nabla} \right)T \right. \nonumber \\
&&\ \ \ \  \left. +\frac{m{\bf u}}{k_{\rm B}T}\cdot
\left(\frac{\partial}{\partial t}+
\frac{{\bf p}}{m}\cdot\bbox{\nabla} \right){\bf v}
+\frac{\bbox{\nabla}U({\bf r},t)}{k_{\rm B}T}\cdot {\bf u}
\right] \tilde f(1+\tilde f).
\label{A1}
\end{eqnarray}
Using the expressions for the density $\tilde n$
given by (16) and the pressure $\tilde P$ in (17), one finds
\begin{eqnarray}
d\tilde n&=&\frac{3\tilde n}{2 T}dT+\frac{\gamma k_{\rm B}T}{z}dz\,, \cr
d\tilde P&=&\frac{5\tilde P}{2T}dT+\frac{\tilde nk_{\rm B}T}{z}dz\,,
\end{eqnarray}
where $\gamma$ is the variable introduced after (22).
One may combine to these equations to obtain
\begin{eqnarray}
\frac{dT}{T}&=&\frac{\sigma_3}{\tilde P}d\tilde P
-\frac{\sigma_4}{\tilde n}d\tilde n, \cr
\frac{dz}{z}&=&
\frac{1}{k_{\rm B}T}\left(\frac{5\sigma_3}{2\gamma}d\tilde n
-\frac{3\sigma_4}{2\tilde n}d\tilde P\right),
\end{eqnarray}
where the thermodynamic functions $\sigma_3$ and $\sigma_4$ are defined by
\begin{equation}
\sigma_3({\bf r},t)=
\frac{\tilde P \gamma}{\frac{5}{2}\tilde P\gamma-\frac{3}{2}\tilde n^2},
\qquad \sigma_4({\bf r},t)=
\frac{\tilde n^2}{\frac{5}{2}\tilde P\gamma-\frac{3}{2}\tilde n^2}.
\end{equation}

Using the lowest-order hydrodynamic equations given in (18), one finds
that the equations in (A3) reduce to
\begin{eqnarray}
\frac{\partial T}{\partial t}&=&-\frac{2}{3}T(\bbox{\nabla}\cdot{\bf v}_n)
-{\bf v}_n\cdot\bbox{\nabla}T+\frac{2T}{3\tilde n}\sigma_1 \Gamma_{12}[\tilde f], \cr
&& \cr
\frac{\partial z}{\partial t}&=&-{\bf v}_n\cdot\bbox{\nabla}z+
\sigma_2z\frac{\Gamma_{12}[\tilde f]}{\tilde n},
\end{eqnarray}
where the local equilibrium thermodynamic functions $\sigma_1$ and $\sigma_2$ are
defined in (22).
The analogous equation for $\partial {\bf v}_n/\partial t$ is given directly
by (18).
Using these results in (A1), one finds that it reduces to
\begin{eqnarray}
&&\left(\frac{\partial}{\partial t}
+\frac{{\bf p}}{m}\cdot\bbox{\nabla}-\bbox{\nabla}U\cdot\bbox{\nabla}_{\bf p}
\right)\tilde f \cr
&&=\left\{
\frac{1}{T}{\bf u}\cdot\bbox{\nabla}T\left(\frac{mu^2}{2k_{\rm B}T}
-\frac{5\tilde P}{2\tilde nk_{\rm B}T} \right)
+\frac{m}{k_{\rm B}T}\left[{\bf u}\cdot({\bf u}\cdot\bbox{\nabla}){\bf v}
-\frac{u^2}{3}\bbox{\nabla}\cdot{\bf v}\right]\right. \cr
&&+\left.\left(\sigma_2+\frac{mu^2}{3k_{\rm B}T}\sigma_1+\frac{m}{k_{\rm B}T}
{\bf u}\cdot{\bf w} \right)\frac{\Gamma_{12}[\tilde f]}{\tilde n}\right\}\tilde f(1+\tilde f),
\end{eqnarray}
where we recall ${\bf u}\equiv {\bf p}/m-{\bf v}_n$.
This can be rewritten in the form shown on the left hand side of
(21).

Using $dz/z=\beta[-(\tilde\mu-U)dT/T+d\tilde\mu-2gdn]$, one obtains from
(A3)
\begin{equation}
d\tilde\mu=\frac{\sigma'_1}{\tilde n}d\tilde P
+\frac{\sigma'_2}{\beta\tilde n}d\tilde n+2g(d\tilde n+d n_c).
\end{equation}
Here $\sigma'_1$ and $\sigma'_2$ are obtained from $\sigma_1$ and
$\sigma_2$ in (22) with the replacement of $\frac{1}{2}m({\bf v}_n-
{\bf v}_c)^2 +\mu_c$ by $\tilde\mu$. The relations (A2)-(A4) and (A7) 
are used in Sec. IV to eliminate $\delta\mu_{\rm diff}$ and $\delta T$ 
from the linearized hydrodynamic equations in (35) and (36).

\section{}
In this Appendix, we sketch the derivation of the approximate
expression for the relaxation rate $\Gamma_{\rm R}$
given in (45).
It is convenient to introduce the five-component vector
\begin{equation}
{\bf y}^{\rm T}=(\delta\bar n,\delta\bar P, \delta\bar v_n, \delta\bar n_c,\delta\bar v_c).
\end{equation}
The linearized hydrodynamic equations in (44) can then be written in the matrix form
\begin{equation}
i\bar\omega{\bf y}=K{\bf y},
\end{equation}
where the $5\times 5$ matrix $K$ is given by
\begin{equation}
K = \left ( \matrix{ \frac{\sigma_{20}\bar n_{c0}}{\bar\tau_{12}\bar n_0} &
\frac{\sigma_{10}\bar n_{c0}}{\bar\tau_{12}\bar n_0} &
\bar n_0\bar k & 
\frac{\lambda\bar n_{c0}}{\bar \tau_{12}t} & 0
\cr
-\frac{2}{3}\left(\frac{\sigma_{20}\lambda\bar n_{c0}^2}{\bar\tau_{12}t\bar n_0}
+\frac{\sigma_{40}\bar\kappa\bar k^2}{\bar n_0}\right)
& 
-\frac{2}{3}\left(\frac{\sigma_{10}\lambda\bar n_{c0}^2}{\bar\tau_{12}t\bar n_0}
-\frac{\sigma_{30}\bar\kappa\bar k^2}{\bar P_0}\right)
& \frac{5}{3}\bar P_0\bar k & -\frac{2\lambda^2\bar n_{c0}^2}{3\bar\tau_{12}
t^2} & 0 \cr
-\frac{6}{5}\lambda\bar k & -\frac{3t}{5\bar n_0}\bar k &
\frac{4\bar\eta}{3\bar n_0}\bar k^2 & -\frac{6}{5}\lambda\bar k & 0 \cr
-\frac{\sigma_{20}\bar n_{c0}}{\bar\tau_{12}\bar n_0} &
-\frac{\sigma_{10}\bar n_{c0}}{\bar \tau_{12}\bar n_0} & 0 &
-\frac{\lambda\bar n_{c0}}{\bar\tau_{12}t} & \bar n_{c0}\bar k \cr
-\frac{6}{5}\lambda\bar k & 0 & 0 & -\frac{3}{5}\lambda\bar k & 0}
 \right)\,.
\end{equation}
Although it is difficult to obtain general analytical solutions of the matrix equation (B2),
we can obtain approximate analytical expression for $\Gamma_{\rm R}$ by making use of the
expression for the determinant of this matrix at zero frequency:
\begin{equation}
{\rm det}K
=\tau_0^5(\Omega_1^2+\Gamma_1^2)(\Omega_2^2+\Gamma_2^2)\Gamma_{\rm R}.
\end{equation}
Working to first order in $\bar\kappa$, $\bar\eta$, $1/\bar \tau_{12}$ 
and second order in $\lambda$, a direct evaluation of the determinant 
of the matrix in (B3) gives
\begin{eqnarray}
{\rm det}K&=&\frac{3}{5}\lambda\bar n_{c0}\frac{t\bar P_0}{\bar n_0}\bar k^4
\left\{\frac{2\sigma_{40}\bar\kappa\bar k^2}{5\bar P_0}
\left(1-\frac{2\lambda\bar n_0^2\sigma_{30}}{t\bar P_0\sigma_{40}}\right) \right. \cr
&& \frac{\bar n_{c0}}{\bar\tau_{12}\bar n_0}\left[\sigma_{20}
+\frac{\lambda\bar n_0}{t\bar P_0}\left(\frac{2}{5}\bar n_{c0}\sigma_{20}
+2\bar P_0\sigma_{10}-2\bar P_0)\right]\right\}.
\end{eqnarray}
If we set $\kappa$, $\eta=0$, the first ($\Omega_1$) and second 
($\Omega_2$) sound mode frequencies for small wavevectors 
are given by the solution of Eq.~(93) of Ref.~\cite{ZNG}.
To second order in $\lambda$, this gives
\begin{equation}
\tau_0^4\Omega_1^2\Omega_2^2=\bar k^4\frac{3}{5}\lambda\bar n_{c0}
\frac{t\bar P_0}{\bar n_0}\left(1-\frac{2\lambda\bar n_0^2}{t\bar 
P_0\sigma_{40}} -\frac{6\lambda\bar n_0^2}{5t\bar P_0}\right).
\end{equation}

Using (B6) and (B5) in (B4), we obtain
\begin{eqnarray}
\Gamma_{\rm R}\tau_0&\approx& \frac{{\rm det}K}{\Omega_1^2\Omega_2^2\tau_0^4} \cr
&=&\frac{\tau_0}{\tau_{\mu}}+\frac{2\bar\kappa\bar k^2}{5\bar P_0}
\left[\sigma_{40}+\frac{\lambda\bar n_0^2}{t\bar P_0}
\left(\frac{6}{5}\sigma_{40}+\frac{2}{5}-2\sigma_{30}\right)\right],
\end{eqnarray}
where $\tau_{\mu}$ is given by
\begin{eqnarray}
\frac{\tau_{12}}{\tau_{\mu}}&=&
\frac{n_{c0}}{k_{\rm B}T_0}\left[
\frac{\frac{5}{2}\tilde P_0+2g\tilde n_0n_{c0}+\frac{2}{3}g^2\gamma_0n_{c0}^2}
{\frac{5}{2}\tilde P_0\gamma_0-\frac{3}{2}\tilde n_0^2}-g\right] \cr
&\simeq& \frac{\bar n_{c0}}{\bar n_0}
\left[\frac{5\bar P_0}{2\bar n_0}+\frac{\lambda}{t}
(2\bar n_{c0}\sigma_{40}-\bar n_0)\right]+O(\lambda^2).
\end{eqnarray}
Neglecting the second small term in the square bracket of (B7), we 
obtain for $\Gamma_{\rm R}$ the simple expression
\begin{equation}
\Gamma_{\rm R}\tau_0=\frac{\tau_0}{\tau_{\mu}}+\frac{2\sigma_{40}\bar\kappa}{5\bar P_0}\bar k^2.
\end{equation}
Numerical calculations show that (B7) is very well approximated by (B9).
Using the definitions of the various dimensionless quantities in (41)-(43), (B9) can be
written in the form given in (45).

\end{appendix}

\clearpage

\centerline{\bf FIGURE CAPTIONS}
\begin{itemize}
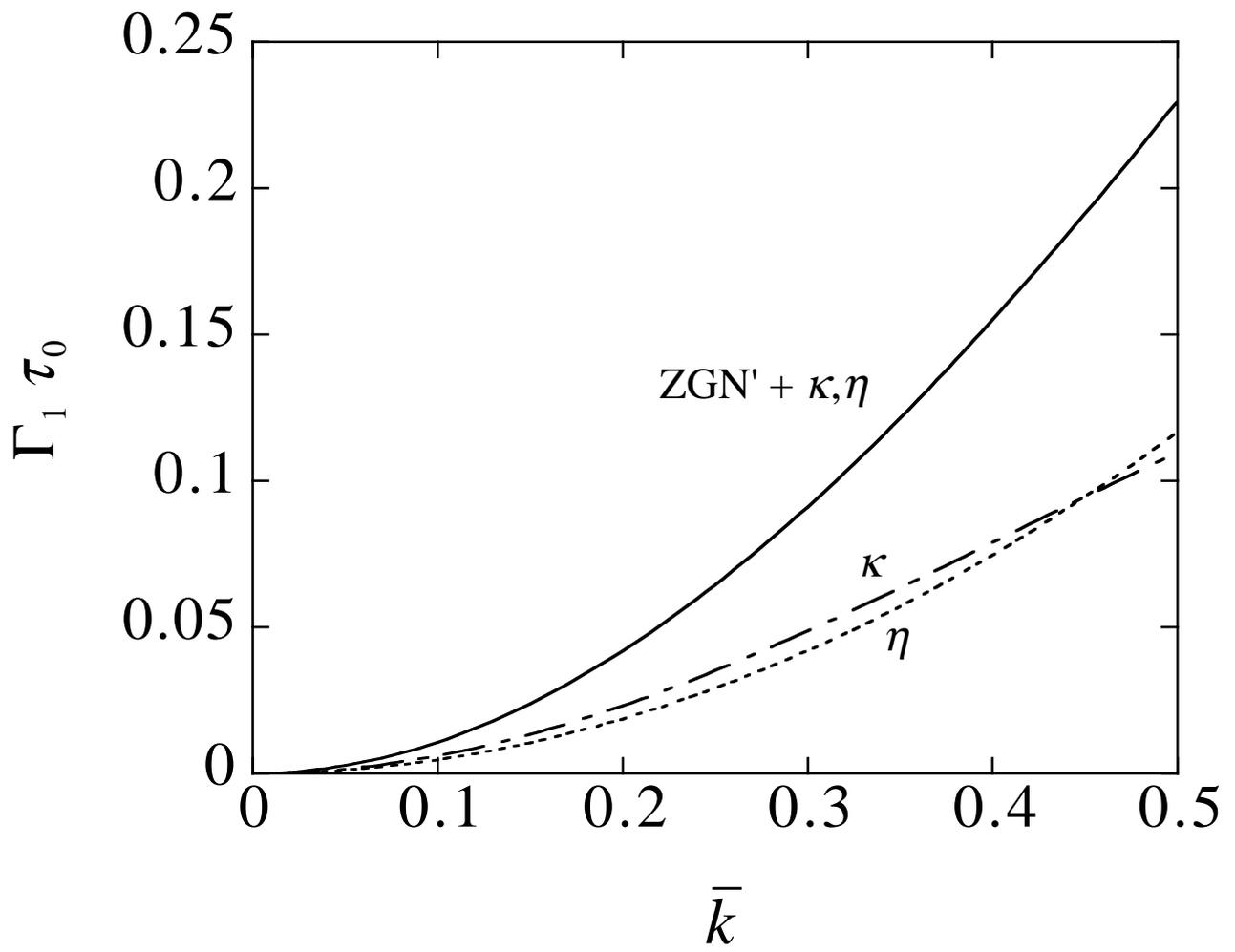
\item[FIG.1:] 
Damping of first sound in a uniform gas for $T/T_{\rm BEC}=0.9$ and 
$gn=0.2k_{\rm B}T_{\rm BEC}$,
as a function of the dimensionless wavevector defined in (42).
The ZGN$'$ results including only $\kappa$ and only $\eta$ are also shown.

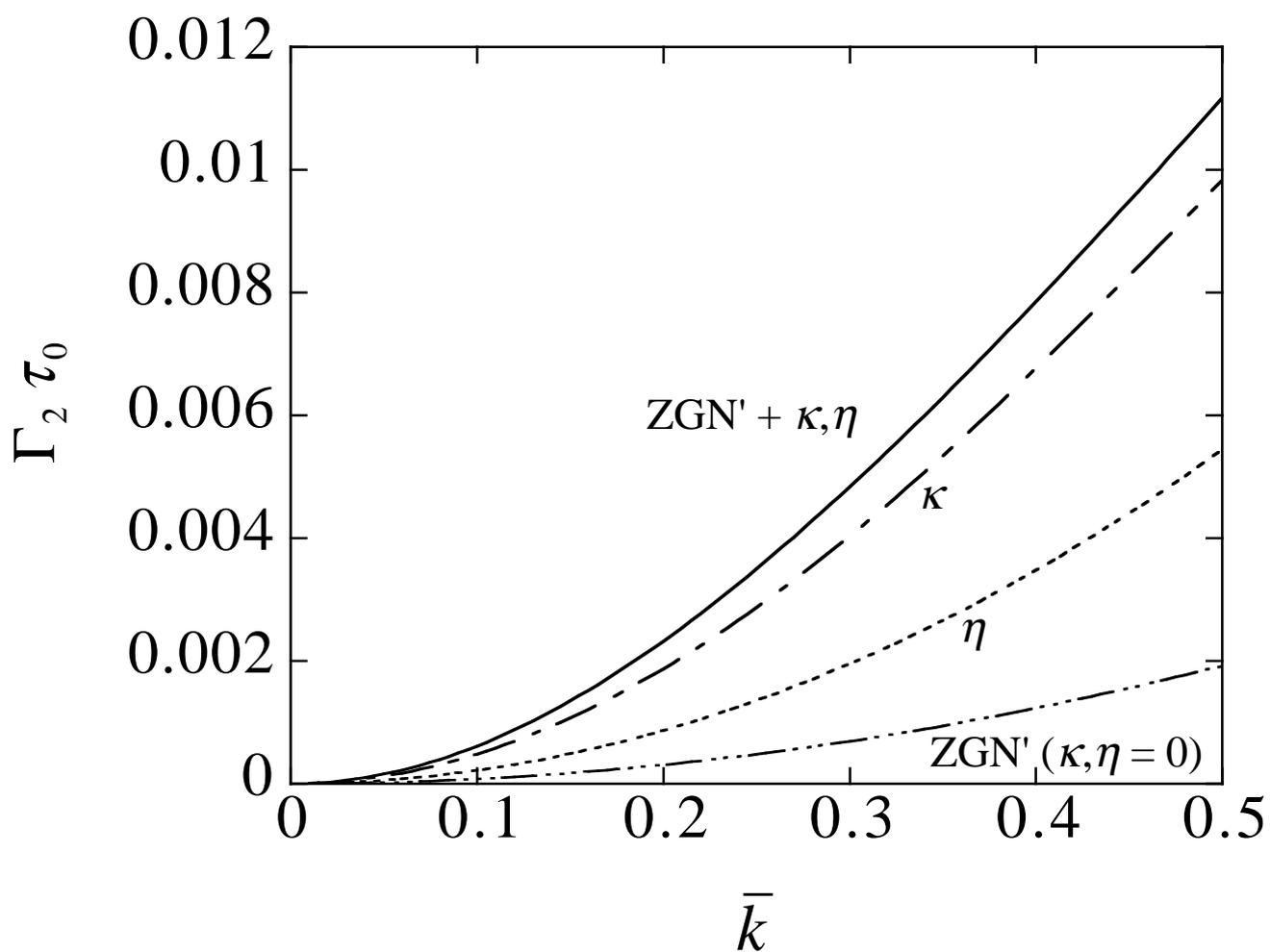
\item[FIG.2:] 
Damping of second sound in a uniform gas for the same parameters as in 
Fig.~1. 

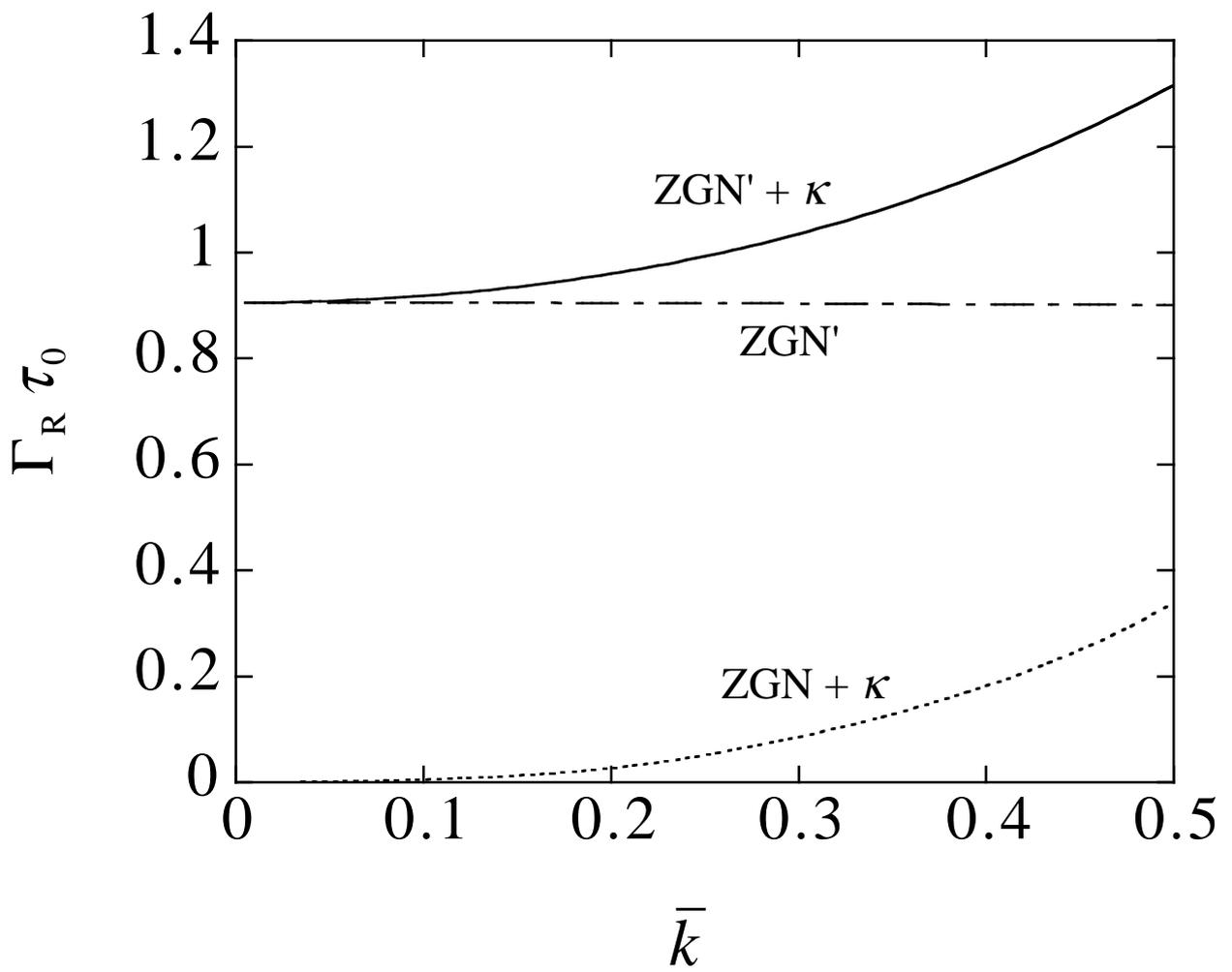
\item[FIG.3:] 
Damping of the relaxational mode vs. wavevector (see Fig.~1).
The damping due to the thermal conductivity is shown for both the 
ZGN ($\tau_{\mu}\to\infty$) and ZGN$'$ theories.
The effect of the shear viscosity is negligible.

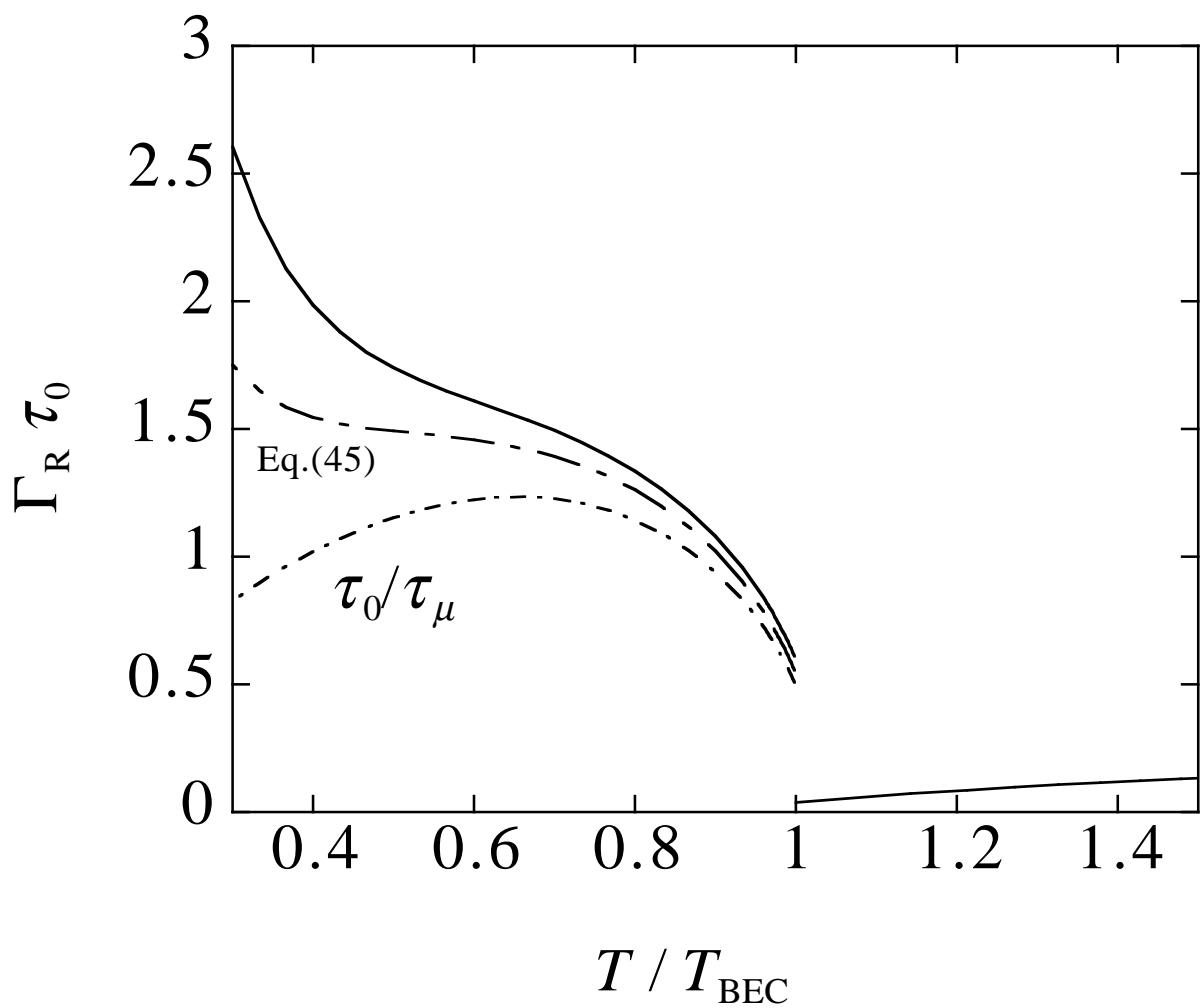
\item[FIG.4:]
Damping of the relaxational mode vs. temperature, for $gn=0.1k_
{\rm B}T_{\rm BEC}$ and $\bar k=0.4$. The ZGN$'$ relaxation time 
$\tau_{\mu}$ is also shown.  All results are normalized to the 
classical gas collision time $\tau_0$ defined below (42).
The approximate analytical expression for $\Gamma_{\rm R}$ in (45) is
compared with a direct numerical solution of the linearized equations 
in (44). The discontinuity at $T_{\rm BEC}$ is spurious, being a result
of the mean-field approximation used for the thermal 
excitations~\cite{NZG}.
\end{itemize}

\begin{references}
\bibitem[*]{griffin}
Permanent address:
Department of Physics, University of Toronto, Toronto,
Ontario, Canada, M5S 1A7.

\bibitem[\dagger]{zaremba}
Permanent address:
Department of Physics, Queen's 
University, Kingston, Ontario, Canada, K7L 3N6.



\bibitem{RMP} F. Dalfovo, S. Giorgini, L.P. Pitaevskii and S. Stringari,
Rev. Mod. Phys. {\bf 71}, 463 (1999).

\bibitem{varenna} See articles in 
{\it Bose-Einstein Condensation in Atomic Gases},
Proceedings of the International School of Physics ``Enrico Fermi'' ,
ed. by M. Inguscio, S. Stringari and C. Wieman (IOS Press, Amsterdam, 1999).

\bibitem{NP} P. Nozi\`eres and D. Pines, {\it The theory of Quantum Liquids}
(Addison-Wesley, Redwood City, California, 1990), Vol. II.

\bibitem{NZG}T. Nikuni, E. Zaremba and A. Griffin,
Phys. Rev. Lett., {\bf 83}, 10 (1999).

\bibitem{ZGN}E. Zaremba, A. Griffin and T. Nikuni,
Phys. Rev. {\bf A57}, 4695 (1998).

\bibitem{ZNG}E. Zaremba, T. Nikuni and A. Griffin,
J. Low Temp. Phys., {\bf 116}, 277 (1999).

\bibitem{Landau} L.D. Landau, J. Phys. (U.S.S.R.) {\bf 5}, 71 (1941).

\bibitem{Khal} I.K. Khalatnikov, {\it An Introduction to the Theory of Superfluidity}
(W. A. Benjamin, New York, 1965).

\bibitem{KD} T.R. Kirkpatrick and J.R. Dorfman, J. Low Temp. Phys.
{\bf 58}, 304 (1985); {\bf 58}, 399 (1985). 

\bibitem{GG} C. Gay and A. Griffin, J. Low Temp. Phys. {\bf 58}, 479 (1985).

\bibitem{NG} T. Nikuni and A. Griffin, J. Low Temp. Phys.
{\bf 793}, 793 (1998).

\bibitem{MIT} D.M. Stamper-Kurn, H.-J. Miesner, S. Inouye,
M.R. Andrews and W. Ketterle, Phys. Rev. Lett. {\bf 81}, 500 (1998).

\bibitem{stoof} H.T.C. Stoof, J. Low Temp. Phys. {\bf 114}, 11 (1999).

\bibitem{huang} K. Huang, {\it Statistical Mechanics} (J. Wiley, New York, 1987),
2nd ed.

\bibitem{UU} E. A. Uehling and G.E. Uhlenbeck, Phys. Rev. {\bf 43}, 552 (1933).
\bibitem{ZNG2} E. Zaremba, T. Nikuni and A. Griffin, to be published.

\bibitem{MIT2} D.M. Stamper-Kurn, A.P. Chikkatur, A. G\"orlitz, S. Inouye, S. Gupta,
D.E. Pritchard and W. Ketterle, Phys. Rev. Lett. {\bf 83}, 2876 (1999).

\bibitem{KPS} G.M. Kavoulakis, C.J. Pethick and H. Smith, Phys. Rev. A {\bf 57}, 2938 (1998).
\end{references}
\end{document}